\documentclass[a4paper]{jpconf}
\usepackage{graphicx}
\usepackage{amsmath,amsfonts}
\usepackage{times}
\usepackage[T1]{fontenc}
\usepackage{epsfig}
\usepackage{amssymb}
\usepackage[english]{babel}
\begin{document}
\title{Supersymmetric extensions of K field theories}

\author{C Adam$^1$, J M Queiruga$^1$, J Sanchez-Guillen$^1$ and A Wereszczynski$^2$}

\address{$^1$ Departamento de F\'isica de Part\'iculas, Universidad
de Santiago, and Instituto Galego de F\'isica de Altas Enerx\'ias
(IGFAE) E-15782 Santiago de Compostela, Spain}

\address{$^2$Institute of Physics,  Jagiellonian University,
Reymonta 4, Krak\'{o}w, Poland}

\ead{adam@fpaxp1.usc.es, queiruga@fpaxp1.usc.es, joaquin@fpaxp1.usc.es, wereszczynski@th.if.uj.edu.pl}

\begin{abstract}
We review the recently developed supersymmetric extensions of field theories with non-standard kinetic terms (so-called K field theories) in two an three dimensions. Further, we study the issue of topological defect formation in these supersymmetric theories. Specifically, we find supersymmetric K field theories which support topological kinks in 1+1 dimensions as well as supersymmetric extensions of the baby Skyrme model for arbitrary nonnegative potentials in 2+1 dimensions.  

\end{abstract}

\section{Introduction}

There exists an intimate relation between topological solitons and supersymmetry. In nonlinear field theories supporting topological solitons there frequently exists a nontrivial lower bound (Bogomolny bound) on the energy in terms of a topological charge (e.g. a winding number) as well as solutions (so-called BPS solutions) saturating this bound. These BPS solutions obey a simpler system of first order equations (the BPS equations) which may be found by expressing the energy functional as a complete square. On the supersymmetry side, it turns out that for the supersymmetric extensions of these theories, the BPS solutions may be understood as a special class of solutions which are invariant under part of the SUSY transformations. Further, the topological charges of the soliton solutions induce central extensions of the SUSY algebra in the sypersymmetrically extended theories. 

The simplest setting where all these related phenomena can be found is the case of a standard real scalar field theory
\begin{equation} \label{stand-lag}
L= \frac{1}{2}\partial_\mu\phi \partial^\mu \phi -V(\phi)
\end{equation}
in 1+1 dimensions,
where the potential $V$ should be non-negative and have at least two vacua $\phi_i  : V(\phi_i)=0$. This theory then generically supports topological kink solutions of finite energy interpolating between two adjacent vacua.  
The energy for static configurations is
\begin{equation}
E= \int dx \left( \frac{1}{2} \phi '^2 + V \right) = \frac{1}{2} \left[ \int dx \left( \phi ' \mp \sqrt{2V} \right)^2 \pm \int dx \sqrt{2V}\phi ' \right]
\end{equation}
from which the BPS equations 
\begin{equation}
\phi'  \mp \sqrt{2V} =0
\end{equation}
and the Bogomolny bound
\begin{equation}
E \ge \pm \int_{-\infty}^\infty \sqrt{2V}\phi ' = \int_{\phi(-\infty)}^{\phi (\infty)} W' (\phi) d\phi = W(\phi(\infty)) - W(\phi(-\infty))
\end{equation} 
for kink and antikink easily follow. Here, the prepotential (or integrating function) $W(\phi)$, which is a function of the scalar field $\phi$, is defined by 
\begin{equation}
W'(\phi) = \sqrt{2V(\phi)}.
\end{equation}

Further,
the standard scalar field theory has a supersymmetric (SUSY) extension based on the superfield
$${\cal S}=- \frac{1}{4}D^{\alpha}\Phi D_{\alpha}\Phi - P(\Phi) $$
 where 
$$\Phi(x,\theta)=\phi(x)+\theta^{\alpha}\psi_{\alpha}(x)-\theta^2 F(x) $$
$$ \quad \quad D_\alpha = \frac{\partial}{\partial\theta^\alpha} -i \gamma^\mu{}_\alpha{}^\beta
          \theta_\beta \partial_\mu \equiv \partial_\alpha+i\theta^\beta\partial_{\alpha\beta} ,$$ 
$\phi$ is a real scalar field, 
$ \alpha =1,2$, $\theta_\alpha $ are the Grassmann-valued superspace coordinates, $\theta^2 \equiv (1/2) \theta^\alpha \theta_\alpha$, $\psi_\alpha $ is a Majorana spinor, and $F$ is the auxiliary field. We use the metric convention $\eta_{\mu\nu} = {\rm diag}(+,-,\ldots ,-)$. Further, the superpotential $P$ is an, in principle arbitrary, function of the superfield. The resulting Lagrangian density in ordinary space is
\begin{equation}
          L=\int d^2 \theta {\cal S} = \frac{1}{2}F^2 - \frac{1}{2}i \psi^\alpha \partial_{\alpha\beta}\psi^\beta +
                      \frac{1}{2}\partial_\mu\phi\partial^\mu \phi  -
                      \frac{1}{2}P''(\phi)\psi^\alpha \psi_\alpha - P'(\phi)F .
\end{equation}
It may be checked easily that after elimination of the auxiliary field with the help of its algebraic field equation the bosonic sector of the theory coincides with the Lagrangian (\ref{stand-lag}) upon identifying $V=(1/2)P'^2=(1/2)W'^2$, which implies $P=W$. In addition, it has been demonstrated long ago \cite{witten-olive}
that the SUSY algebra of this theory contains central extensions which are related to the topological charge of the kink, and that a certain linear combination of the supercharges of the theory is proportional to the BPS equation of the kink, which implies the invariance of the kink solution under the same linear combination of SUSY transformations.

If one wants to carry over these relations between topological solitons and SUSY to higher dimensions then some additional structure has to be introduced, because the simple scalar field theory (\ref{stand-lag}) cannot have static finite energy solutions, as a consequence of Derrick's theorem. A wellknown solution to this problem consists in the introduction of gauge fields. Concretely, one assumes several scalar fields ("Higgs fields") instead of just one and couples them to abelian or non-abelian gauge fields via the usual minimal coupling. The resulting theories then evade the Derrick theorem and may support topological solitons. Specifically, 
the abelian Higgs (or Chern--Simons Higgs) models in 2+1 dimensions (vortices), the theory of t´Hooft-Polyakov in 3+1 dimensions (monopoles), or pure Yang--Mills theory in 4+0 dimensions (instantons) all support topological solitons which are, at the same time, BPS solutions saturating a Bogomolny bound. On the other hand, all these theories are well-known to allow for supersymmetric extensions \cite{divecchia-ferrara}, \cite{witten-olive}, \cite{edelstein-nunez}, \cite{d'adda-horsley}, \cite{d'adda-divecchia}, 
where, in addition, the topological charges are related to central extensions in the correspondiong SUSY algebras, and the BPS soliton solutions are invariant w.r.t. part of the SUSY transformations. In the higher-dimensional cases this usually requires an extension to higher (e.g. $N=2$) supersymmetry. We remark that these supersymmetric theories have
been studied intensively over the last decades, and the corresponding literature is too extensive to be quoted here. 

There exists, however, a second possibility to evade the Derrick theorem and construct field theories supporting topological solitons in higher dimensions, which consists in allowing for non-standard kinetic terms (so-called K field theories), usually higher (than second) powers of first derivatives in the Lagrangian.
Probably the most famous model of this type which allows for topological solitons is the Skyrme model \cite{skyrme} in 3+1 dimensions with the fields taking values in the group SU(2). Other wellknown models are the so-called Skyrme--Faddeev--Niemi (SFN) model (a $S^2$ (or CP(1)) restriction of the Skyrme model in 3+1 dimensions) \cite{SFN} or the baby Skyrme model in 2+1 dimensions \cite{piette1} - \cite{weidig}. All three  models support topological solitons which may be classified by a topological index (a winding number for the Skyrme and baby Skyrme model, a linking number - the Hopf index - for the SFN model). Further, there exists a Bogomolny bound in all three models, but nontrivial soliton solutions do not saturate this bound (that is to say, they are not BPS solutions), exemplifying in this way already one major difference with the theories with standard kinetic terms described in the last paragraph. We want to emphasize, however, that both the Skyrme \cite{BPSskyrme} and the baby Skyrme models \cite{comp-bS} have certain restricted submodels where nontrivial BPS soliton solutions do exist. 

Much less is known about possible supersymmetric extensions of this second type of topological soliton models.
Supersymmetric extensions of the SFN model were investigated in \cite{nepo} and in \cite{frey}. In both papers, a formulation of the SFN model was used where the CP(1) restriction of the Skyrme model is achieved via a gauging of the third, unwanted degree of freedom. As a result, the SFN model is expressed by two complex scalar fields and an undynamical gauge field, which are then promoted to two chiral superfields and a real vector superfield in the Wess--Zumino gauge, respectively. 
The result of the analysis is that the SFN model as it stands cannot be supersymmetrically extended by these methods. Instead, the supersymmetric extension contains additional terms already in the bosonic sector, and also the field equations of the bosonic fields are different.
In \cite{frey} also the issue of Bogomolny type bounds of the supersymmetrically extended SFN type models was investigated. It results that the $N=1$ supersymmetrically extended model has a Vakulenko--Kapitansky \cite{KaVa} type bound $M_s \ge c Q_H^{3/4}$ like the original, nonsupersymmetric SFN model, despite the additional terms in the action (here $M_s$ denotes a soliton mass, $Q_H$ is the  Hopf index and $c$ is a known constant). On the other hand, the $N=2$ SUSY extension has a bound linear in the corresponding topological index, and this bound may be calculated from the central extensions of the SUSY algebra of the model. The topological index is still a Hopf index, but it slightly differs from the Hopf index of the original and $N=1$ SUSY SFN models, because it depends both on the original CP(1) field and an auxiliary field.

Quite recently, we found that the baby Skyrme models in 2+1 dimensions allow for $N=1$ supersymmetric extensions for arbitrary nonnegative potentials \cite{bS}. Here it came as a certain surprise that precisely the restricted submodels where nontrivial BPS soliton solutions exist and may be found analytically are the ones that {\em cannot} be supersymmetrically extended, at least not by the methods developed and used in \cite{bS}.

In a different line of development, scalar field theories with a non-standard kinetic term (i.e., K field theories) have been studied with increasing effort during the last years, beginning with the observation about a decade ago of their possible relevance for the solution of some problems in cosmology (k-inflation
\cite{k-infl} and k-essence \cite{k-ess}).
Further applications of K fields to cosmological issues may be found, e.g., in \cite{A-H1} - \cite{liu}, whereas other, more formal or mathematical aspects of K field theories, like the existence of topological defects with compact support (so-called compactons) have been studied, e.g., in \cite{werle} -\cite{fring}. Well-posedness of the K field system and the issue of signal propagation in K field backgrounds has been investigated, e.g., in \cite{bab-muk-1} and, recently, in \cite{bergli1}. Depending on their vacuum structure,
scalar K field theories may support topological solitons (e.g., kinks), like their standard counterparts, see e.g. \cite{babichev1}, \cite{bazeia3}, \cite{comp}. So the question naturally arises whether these K field theories allow for supersymmetric extensions, and whether the tight relation between BPS equations and topological charges on the one hand, and the SUSY algebra and its central extensions on the other hand may be maintained also for K field theories. If K field theories turn out to have applications for the solution of cosmological problems in the very early universe, where SUSY is assumed unbroken, then these possible supersymmetric extensions may even be relevant from a phenomenological perspective. 

The investigation of SUSY extensions of scalar K field theories has been resumed very recently. Concretely, 
\cite{bazeia2}, 
\cite{susy2} and \cite{twins} studied supersymmetric extensions of K field theories in 1+1 dimensions, based on a SUSY representation which may be employed both in 1+1 and in 2+1 dimensions. On the other hand,  the investigations of \cite{ovrut1} and \cite{ovrut2} were for 3+1 dimensional K theories, and with some concrete cosmological applications (ghost condensates and Galileons) in mind. 
All classes of SUSY K field theories in 1+1 dimensions studied in \cite{bazeia2}, 
\cite{susy2} and \cite{twins} contain theories which support topological solitons (kinks) in their bosonic sector. Further, it remains true for all of them that the static Euler--Lagrange equations in the bosonic sector may be integrated once to first order equations. On the other hand, not all of these first order equations are, in fact, BPS equations. Apparently, genuine BPS equations only result if the auxiliary field $F$ shows up at most quadratically in the SUSY Lagrangian, although this is a mere conjecture at the moment, based on the investigation of specific models. The issues of central extensions of the SUSY algebras and the relation between SUSY transformations and BPS equations for supersymmetric K field theories are open problems at the moment, mainly due to the complicated structure of the fermionic sector of SUSY K field theories.

It is the main purpose of the present contribution to review the results of \cite{susy2} (and to a certain degree also of \cite{bazeia2}) for SUSY K field theories in 1+1 dimensions as well as the results of \cite{bS} on the SUSY baby Skyrme models. We also want to shed some light on similarities and differences with the case of SUSY extensions of standard field theories, describe open problems and point towards relevant and interesting future research directions in this novel subject. Our contribution is organized as follows. In Section 2 we first introduce a set of superfields which we want to use as "building blocks" for the construction of SUSY Lagrangians. Then we study a class of SUSY Lagrangians first found in \cite{susy2} which, in the bosonic sector, may still be written as a sum of a generalized kinetic term and a potential. These Lagrangians may give rise to topological kink solutions, and we provide some concrete examples. On the other hand, these Lagrangians are not of the BPS type. Then, in Section 2.3, we discuss another class of SUSY Lagrangians originally found in \cite{bazeia2}. They cannot be written as a kinetic term plus a potential, but nevertheless may provide topological kink solutions. Further, these Lagrangians are at most quadratic in the auxiliary field, and they have the BPS property. 
In Section 3, we briefly decribe the SUSY extensions of baby Skyrme models originally found in \cite{bS}. Section 4 contains a discussion of the results presented in this contribution. 

\section{Supersymmetric Lagrangians}

For the Lagrangians we want to construct we shall need the following superfields 
\begin{equation}\label{kin-SF1}
D^\alpha \Phi D_\alpha \Phi  = 2\psi^2 -2\theta^\alpha (\psi_\alpha F + i  \psi^\beta \partial_{\alpha\beta} \phi )
 + 2\theta^2 (F^2 -i\psi^\alpha \partial_{\alpha\beta} \psi^\beta + \partial_\mu \phi \partial^\mu \phi )
\end{equation}
\begin{eqnarray}
D^\beta D^\alpha \Phi D_\beta D_\alpha \Phi &=& 2\left( \partial_\mu \phi \partial^\mu \phi -\theta^\gamma \partial^{\alpha\beta} \psi_\gamma \partial_{\alpha\beta} \phi + \theta^2 \partial^{\alpha\beta} \phi \partial_{\alpha\beta}F + \frac{1}{2} \theta^2 \partial^{\alpha\beta} \psi^\gamma
\partial_{\alpha\beta} \psi_\gamma \right. \nonumber \\
&& \left. + F^2 -2i F \theta^\gamma \partial_{\delta\gamma} \psi^\delta +2\theta^2 F\Box \phi + \theta^2 \partial_\delta{}^\gamma \psi^\delta \partial_{\beta\gamma}\psi^\beta \right) \label{kin-SF2}
\end{eqnarray}
\begin{equation} \label{kin-SF3}
D^2 \Phi D^2 \Phi =  F^2 -2i F \theta^\gamma \partial_{\delta\gamma} \psi^\delta +2\theta^2 F\Box \phi + \theta^2 \partial_\delta{}^\gamma \psi^\delta \partial_{\beta\gamma}\psi^\beta 
\end{equation}
as well as their purely bosonic parts 
\begin{eqnarray}
(D^\alpha \Phi D_\alpha \Phi)_{\psi =0} &=& 2 \theta^2 (F^2 + \partial^\mu \phi \partial_\mu \phi )  \label{D-eq-1}\\
(D^\beta D^\alpha \Phi D_\beta D_\alpha \Phi)_{\psi =0} &=& 2(F^2 + \partial^\mu \phi \partial_\mu \phi )  +
 4\theta^2 (F \Box \phi  - \partial_\mu \phi \partial^\mu F  )  
\label{D-eq-2} \\
(D^2\Phi D^2 \Phi)_{\psi =0} &=& F^2 +2 \theta^2 F \Box \phi  . \label{D-eq-3}
\end{eqnarray}
We remark that all spinorial contributions to the Lagrangians we consider are at least quadratic in the spinors, therefore it is a consistent restriction to study the subsector with $\psi_\alpha =0$. We shall, in fact, restrict to the bosonic sector in the following, because we are mainly interested in bosonic topological soliton solutions in the SUSY theories we construct below. 
Next, we have to find the supersymmetric actions we want to investigate. A supersymmetric action always is the superspace integral of a superfield. Further, due to the Grassmann integration rules $\int d^2\theta  =0$, $\int d^2\theta \theta_\alpha =0$, $\int d^2\theta \theta^2 =-1=D^2 \theta^2$, the corresponding Lagrangian in ordinary space-time always is the $\theta^2$ component of the superfield. 
Concretely, we want to use the following Lagrangians as building blocks
\begin{eqnarray}
( {\cal L}^{(k,n)})_{\psi =0} &=& \Bigl(  [(\frac{1}{2} D^\alpha \Phi D_\alpha \Phi) (\frac{1}{2} D^\beta D^\alpha \Phi
D_\beta D_\alpha \Phi )^{k-1}  \; (D^2 \Phi D^2 \Phi )^n ] \Bigr)_{\theta^2 ;\psi =0} \nonumber \\
&=& (F^2 +\partial_\mu \phi \partial^\mu \phi )^k F^{2n} \label{building-b}
\end{eqnarray}
where $k=1,2,\ldots$ and $n=0,1,2,\ldots$ and
the subindex $\theta^2$ just means the $\theta^2$ component of the corresponding superfield.
The idea now is to choose certain linear combinations of the ${\cal L}^{(k,n)}$ with specific properties. In this section we want to choose linear combinations such that the terms where the auxiliary field $F$ couples to the kinetic term $\partial_\mu \phi \partial^\mu \phi$ cancel. The right combination is 
\begin{eqnarray}
( {\cal L}^{(k)})_{\psi =0} &\equiv & 
\sum_{i=0}^{k-1} \binom{k}{i}(-1)^i ( {\cal L}^{(k-i,i)})_{\psi =0} \nonumber \\ 
&=&  (\partial^\mu\phi\partial_\mu\phi)^k + (-1)^{k-1}F^{2k} .
\label{k-n-sum}
\end{eqnarray}
The Lagrangians we want to consider are linear combinations of this expression, where we also include a potential term, because we are mainly interested in topological solitons. That is to say, we add a superpotential $-P(\Phi)$ to the action density in superspace which, in ordinary space-time, induces the bosonic Lagrangian density $ P_{\theta^2} =-P'(\phi) F$ (here the prime denotes the derivative w.r.t. the argument $\phi$). Therefore, the class of Lagrangians we consider is
\begin{eqnarray}
{\cal L}_b^{(\alpha ,P)} &=& \sum_{k=1}^N \alpha_k ({\cal L}^{(k)})_{\psi =0} - P'  F \nonumber \\
&=& \sum_{k=1}^N \alpha_k [ (\partial^\mu\phi\partial_\mu\phi)^k + (-1)^{k-1}F^{2k}] - P'(\phi) F
\end{eqnarray}
where the lower index $b$ means "bosonic" (we only consider the bosonic sector of a supersymmetric Lagrangian), and the upper index $\alpha$ should be understood as a multiindex $\alpha = (\alpha_1 ,\alpha_2 , \ldots , \alpha_N )$ of coupling constants. Further, $N$ is a positive integer. In a next step, we try to eliminate $F$ via its algebraic field equation
\begin{equation} \label{F-eq}
\sum_{k=1}^N (-1)^{k-1} 2k \alpha_k F^{2k-1} - P' (\phi )=0.
\end{equation}
The resulting algebraic equation for $F$ has, in general, no closed solutions. There exists, however, a second, equivalent way to interpret this equation. We  made, in fact, no assumption about the functional form of $P(\phi)$, therefore we may, instead, assume that $F$ is a given but arbitrary function of $\phi$ which in turn determines $P' (\phi)$ via Eq. (\ref{F-eq}). Doing this and eliminating the resulting $P'$ we finally get the 
Lagrangian density
\begin{eqnarray} \label{L-b-F}
{\cal L}_b^{(\alpha ,F)}
&=& \sum_{k=1}^N \alpha_k [ (\partial^\mu\phi\partial_\mu\phi)^k - (-1)^{k-1}(2k-1) F^{2k}] 
\end{eqnarray}
where now $F=F(\phi)$ is a given function of $\phi$ which we may choose freely depending on the theory or physical problem under consideration. 
 
{\em Remark:} By an adequate choice of the constants $\alpha_k$, Lagrangians which obey the required stability criteria (like energy positivity, the null energy condition, etc.) can easily be found \cite{susy2}.

{\em Remark:} For field dependent coefficients $\alpha_k = \alpha_k (\phi)$ the above Lagrangians are still the bosonic sectors of supersymmetric Lagrangians \cite{susy2}. We will, however, assume constant $\alpha_k$ in this article.

{\em Remark:} Up to now, the construction of SUSY Lagrangians is valid both for 1+1 and for 2+1 dimensions. This will be relevant later on in the construction of SUSY baby Skyrme models. 

\subsection{Topological kink solutions}

In the remainder of this section we will, however, discuss static solutions (topological kinks) in 1+1 dimensions. 
For a general Lagrangian ${\cal L}(X,\phi)$ where $X\equiv \frac{1}{2} \partial_\mu \phi \partial^\mu \phi = \frac{1}{2}(\dot{\phi}^2 - \phi '^2)$,
the Euler--Lagrange equation is
\begin{equation}
\partial_\mu ( {\cal L}_{,X} \partial^\mu \phi ) - {\cal L}_{,\phi} =0.
\end{equation}
 Further, the energy momentum tensor reads
\begin{equation}
T_{\mu\nu} = {\cal L}_{,X} \partial_\mu \phi \partial_\nu \phi - g_{\mu\nu} {\cal L} .
\end{equation}
For static configurations $\phi = \phi (x)$, $\phi ' \equiv \partial_x \phi$, the nonzero components of the energy momentum tensor are
\begin{eqnarray} \label{stat-en-de}
T_{00} &=& {\cal E} = -{\cal L} \\
T_{11}  &=& {\cal P} = {\cal L}_{,X} \phi'^2 + {\cal L}
\end{eqnarray}
where ${\cal E}$ is the energy density and ${\cal P}$ is the pressure. Further, for static configurations the Euler--Lagrange equation 
may be integrated once to give
\begin{equation} \label{zero-press}
-2X{\cal L}_{,X} + {\cal L} = \phi '^2 {\cal L}_{,X} + {\cal L} \equiv {\cal P} =0
\end{equation}
(in general, there may be an arbitrary, nonzero integration constant at the r.h.s. of Eq. (\ref{zero-press}), but the condition that the vacuum has zero energy density sets this constant equal to zero).

For the Lagrangian (\ref{L-b-F}), we, therefore, get the once integrated static field equation
\begin{equation} \label{Lb-first-order-eq}
\sum_{k=1}^N (2k-1) \alpha_k \left( -(2X)^k + (-1)^k F^{2k} \right) =0
\end{equation}
which is obviously solved by solutions to the simpler equation
\begin{equation} \label{indep-F-eq}
-2X \equiv \phi'^2 = F^2 \quad \Rightarrow \quad \phi' = \pm F .
\end{equation}
This equation and its solutions only depend on the choice for $F(\phi)$, whereas they do not depend on the Lagrangian (i.e., on the $\alpha_k$). We, therefore, call them generic solutions. Depending on the choice for $F(\phi)$, these generic solutions may be topological solitons. E.g. for the simple choice $F=1-\phi^2$, the solution of  (\ref{indep-F-eq}) is just the well-known $\phi^4$ kink solution
$\phi (x) = \tanh (x-x_0)$, for arbitrary $\alpha_k$.

In addition to the generic solutions, Eq. (\ref{Lb-first-order-eq}) in general has further roots
\begin{equation} \label{spec-eq}
-2X = f_i^2(\phi) \quad \Rightarrow \quad \phi' = \pm f_i(\phi) \; ,\quad i=1,\ldots ,N-1
\end{equation}
and the corresponding solutions (which we shall call "specific solutions"), which now depend on the $\alpha_k$, that is, on the specific Lagrangian in question. Depending on the choice for $F$ and $\alpha_k$, some of these further static solutions may be solitons, too. There also exists the possibility to join different solutions, forming solitons in the class ${\cal C}^1$ of continuous functions with a continuous first derivative, as we shall see in the examples presented below.  

Next, let us briefly discuss the energy of a generic solution, where we choose the  kink
(the solution of $\phi ' = + F$) for concreteness.  We find for the energy density
\begin{eqnarray}
{\cal E} &=& \sum_{k=1}^N (-1)^{k-1} \alpha_k \left(\phi '^{2k} + (2k-1) F^{2k}\right) = \sum_{k=1}^N (-1)^{k-1} \alpha_k 2k F^{2k} 
\nonumber \\
&=& 
\phi ' \sum_{k=1}^N (-1)^{k-1} \alpha_k 2k F^{2k-1} \equiv \phi ' W_{,\phi}
\end{eqnarray}
where we used the kink equation $\phi' =F$ and introduced the prepotential (or integrating function) $W (\phi)$ which must be understood as a function of $\phi$. For the energy this leads to
\begin{equation}
E = \int_{-\infty}^\infty dx \phi ' W_{,\phi} = \int_{\phi (-\infty)}^{\phi (\infty )} d\phi W_{,\phi} = W(\phi (\infty)) - W(\phi (-\infty)),
\end{equation}
 and all that is needed for the evaluation of this energy is the root $\phi ' = F(\phi)$ and the asymptotic behaviour $\phi (\pm\infty)$ of the kink. We remark that the integrating function $W(\phi)$ is identical to the superpotential $P(\phi)$
\begin{equation}
W(\phi)=P(\phi)
\end{equation}
as is obvious from Eq. (\ref{F-eq}). This is exactly as in the case of the standard supersymmetric scalar field theory with the standard, quadratic kinetic term. There remains, however, an important difference. For the standard supersymmetric scalar field theory it is, in fact, possible to rewrite the energy functional for static field configurations in a BPS form, such that both the first order field equations for static fields and the simple, topological expressions $E=P(\phi(\infty))-P(\phi(-\infty))$ for the resulting energies are a consequence of the BPS property of the energy functional.  On the contrary, for the models introduced in this section there is no obvious way to rewrite them in a BPS form,
despite the existence of the first integral (\ref{zero-press}), because the energy functional contains, in general, more than two terms (just two terms are needed to complete a square and arrive at the BPS form). 

We remark that for the additional, specific solutions (solutions of Eq. (\ref{spec-eq})), the relation $W=P$  is no longer true, although it is still possible to calculate the energies of the specific solutions with the help of an integrating function $W(\phi)$.    

\subsection{Examples}

Here, we shall briefly discuss kink solutions for two explicit examples of Lagrangians, where for both examples we choose 
$\alpha_3 = \frac{1}{5}$, $\alpha_2 = \frac{2}{3}$, and $\alpha_1 = 1$, which gives rise to the class of Lagrangian densities
\begin{equation} \label{alpha-lag}
{\cal L}_b^{ex} = \frac{1}{5} (\partial_\mu \phi \partial^\mu \phi )^3 + \frac{2}{3}  (\partial_\mu \phi \partial^\mu \phi )^2 + \partial_\mu \phi \partial^\mu \phi  
- F^6 + 2 F^4 - F^2 
\end{equation}
where the potential term already in terms of $F$ factorizes, $F^6 - 2 F^4 + F^2 = F^2 (1-F^2)^2$, which allows us to easily find potentials with several vacua.

For the first example we choose $F(\phi)=\phi$ with the resulting Lagrangian 
\begin{equation} 
{\cal L}_b^{ex1} = \frac{1}{5} (\partial_\mu \phi \partial^\mu \phi )^3 + \frac{2}{3}  (\partial_\mu \phi \partial^\mu \phi )^2 + \partial_\mu \phi \partial^\mu \phi  
- \phi^2 (1-\phi^2)^2 
\end{equation}
such that the potential has the three minima $\phi = 0,\pm 1$. 
The once-integrated field equation for static solutions reads 
\begin{equation} \label{example-eq}
\phi '^6 -2\phi '^4 + \phi '^2 = \phi^6 -2 \phi^4 + \phi^2 
\end{equation}
and has six roots, namely the obvious generic ones
\begin{equation}
\phi' = \pm \phi \quad \Rightarrow \quad \phi = \exp \pm (x-x_0)
\end{equation}
and the four specific ones
\begin{equation} \label{phi-prime-eq}
\phi ' = \pm \frac{1}{2} \left( \phi \pm \sqrt{ 4-3\phi^2}\right) .
\end{equation}
It turns out that in this example no root provides a solution which interpolates between two vacua (i.e., a kink). Instead, all solutions start at one vacuum and then either reach infinity or hit a singularity. It is, however, possible to join different solutions such that the resulting configuration interpolates between different vacua and is continuous with a continuous first derivative at the joining point. That is to say, it is a kink solution in the class ${\cal C}^1$ of continuous functions with a continuous first derivative. In Figure 1 we display all six static solutions for the "initial" condition $\phi (0)=1/\sqrt{3}$. In Figures 2 and 3 we show two kink solutions which result from the joining procedure. For a more detailed discussion we refer to \cite{susy2}.

\begin{figure}[h!]
\includegraphics[angle=0,width=0.45 \textwidth]{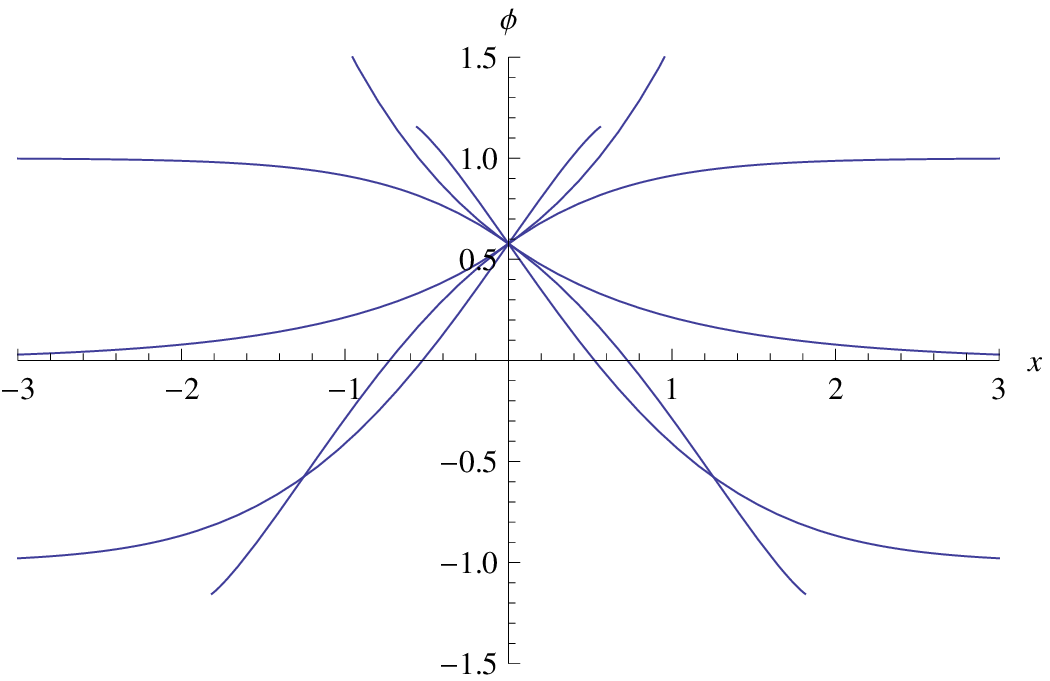}
\includegraphics[angle=0,width=0.45 \textwidth]{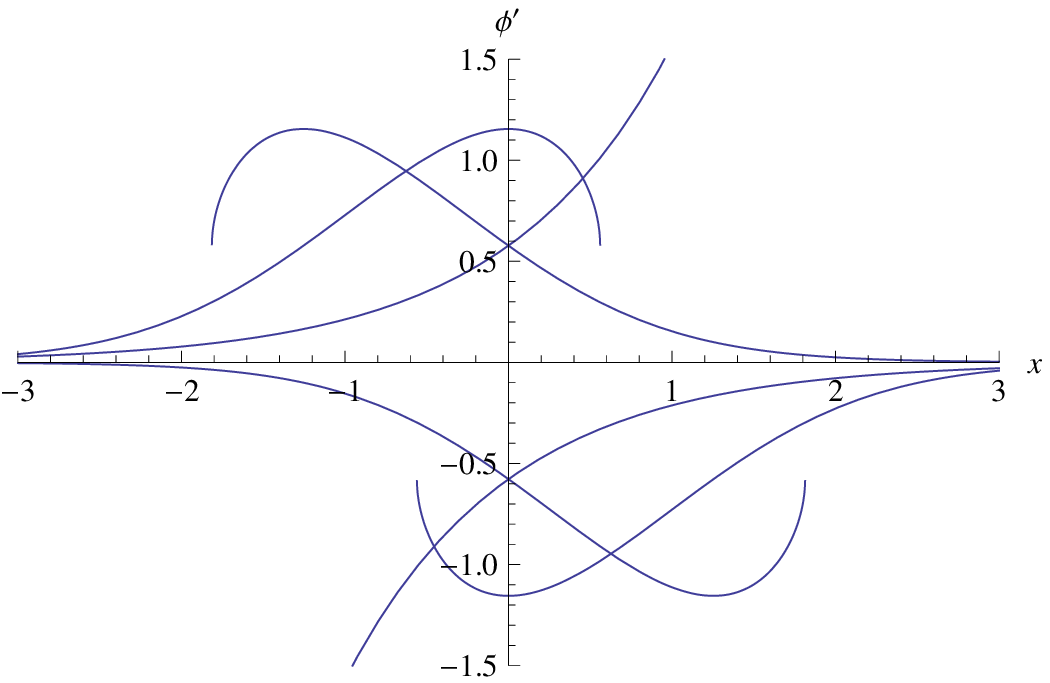}
\caption{ The six solutions of Eq. (\ref{example-eq}) (left figure) and their derivatives (right figure). }\label{fig1}
\end{figure}
\begin{figure}[h!]
\includegraphics[angle=0,width=0.45 \textwidth]{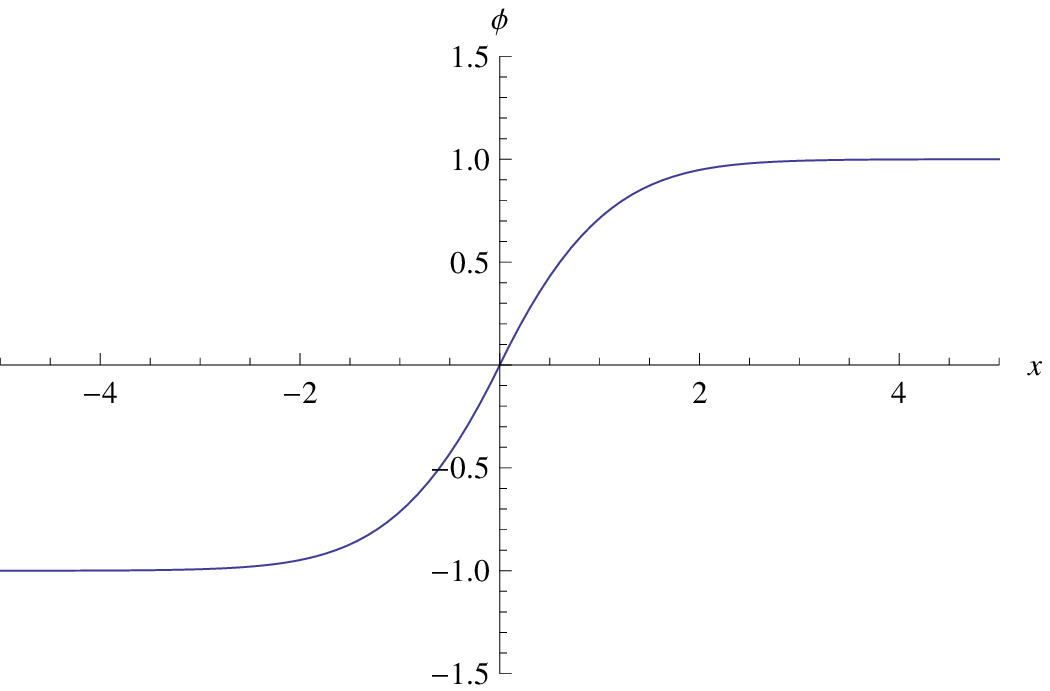}
\includegraphics[angle=0,width=0.45 \textwidth]{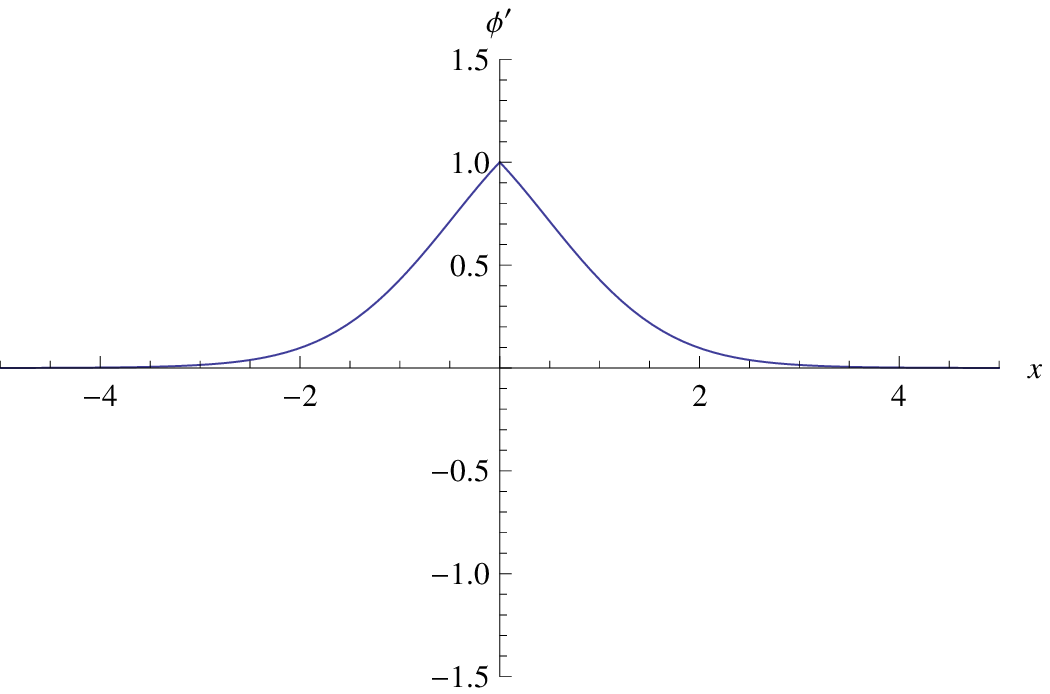}
\caption{The kink interpolating between $-1$ and $1$ and its first derivative.  }\label{fig2}
\end{figure}

\begin{figure}[h!]
\includegraphics[angle=0,width=0.45 \textwidth]{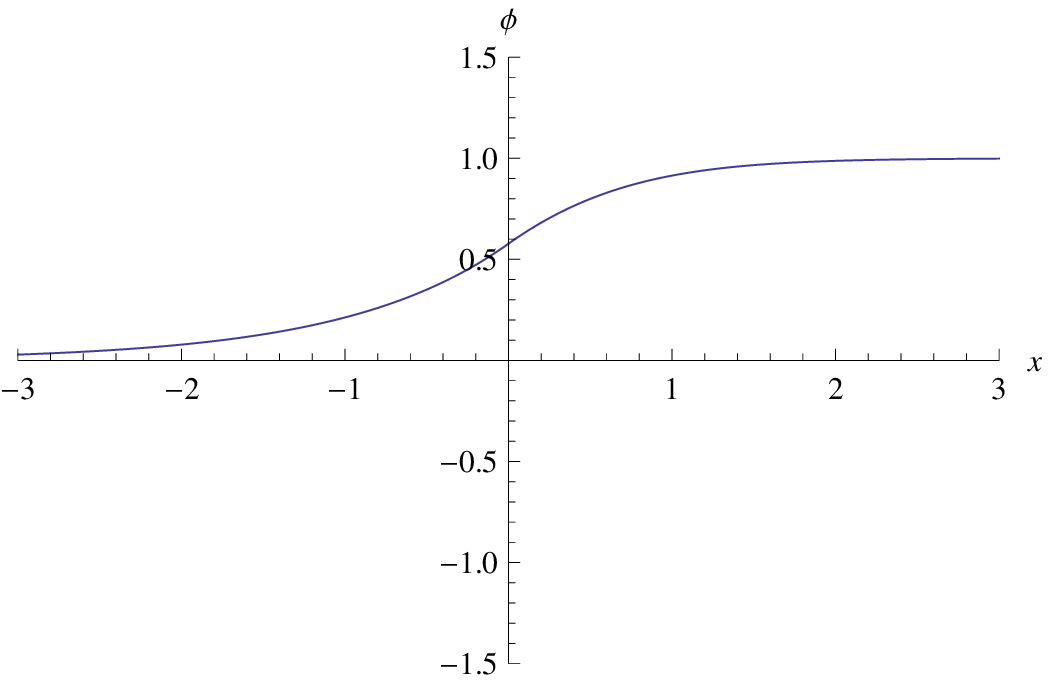}
\includegraphics[angle=0,width=0.45 \textwidth]{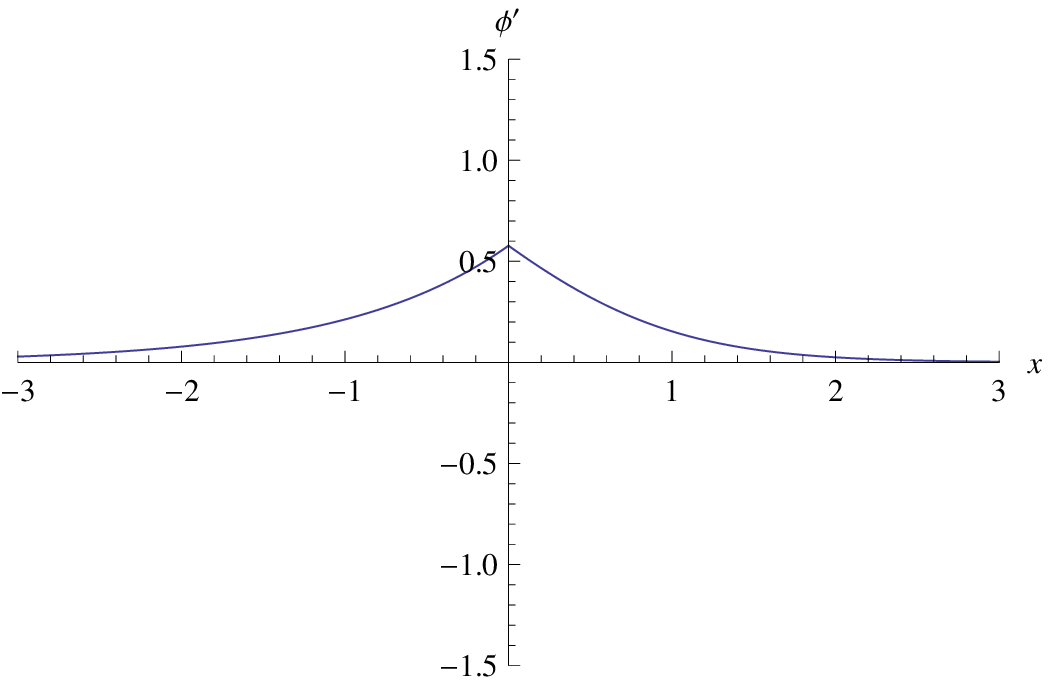}
\caption{The kink interpolating between 0 and 1 and its first derivative.  }\label{fig3}
\end{figure}

For the second example we choose $F=1-\phi^2$ which leads to the Lagrangian
\begin{equation} 
{\cal L}_b^{ex2} = \frac{1}{5} (\partial_\mu \phi \partial^\mu \phi )^3 + \frac{2}{3}  (\partial_\mu \phi \partial^\mu \phi )^2 + \partial_\mu \phi \partial^\mu \phi  
- \phi^4 (1-\phi^2)^2 (2-\phi^2)^2
\end{equation}
and the resulting potential has the five vacua $\phi = 0, \pm 1, \pm \sqrt{2}$. The once integrate static field equation
\begin{equation} \label{V-tildetilde-eq}
\phi'^{6}-2\phi'^4+\phi'^2=(1-\phi^2)^6-2(1-\phi^2)^4+(1-\phi^2)^2 =\phi^4(1-\phi^2)^2(2-\phi^2)^2 
\end{equation}
has again six roots, namely the two generic ones
\begin{equation}
\phi' = \pm (1-\phi^2 )
\end{equation}
which are solved by the $\phi^4$ (anti)kink $\phi = \pm \tanh (x-x_0)$, and the four further specific roots
\begin{eqnarray}
\phi' &=& \pm \frac{1}{2} \left( 1-\phi^2 + \sqrt{1+6\phi^2-3\phi^4} \right) \label{2-2-kink} \\
\phi' &=& \pm \frac{1}{2} \left( 1-\phi^2 - \sqrt{1+6\phi^2-3\phi^4} \right) . \label{0-0-kink}
\end{eqnarray}
This second example has kinks which belong to the ${\cal C}^\infty$ class of functions, namely the generic solutions (the standard $\phi^4$ kink interpolating between the vacua $-1$ and $1$), and the solutions to Eq. (\ref{2-2-kink}) which describe a kink (antikink) interpolating between the vacua $-\sqrt{2}$ and $\sqrt{2}$, see Figure 4. Further kinks in the class of ${\cal C}^1$ functions may be formed by the joining procedure like in example 1. For a detailed discussion we refer, again, to \cite{susy2}.

\begin{figure}[h!]
\includegraphics[angle=0,width=0.45 \textwidth]{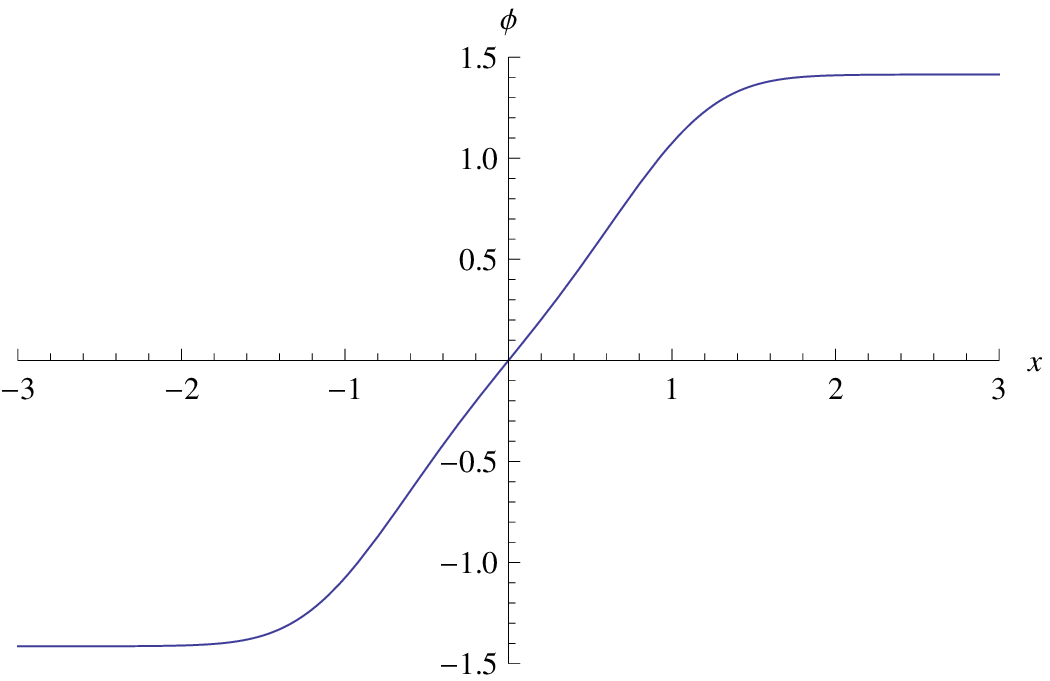}
\includegraphics[angle=0,width=0.45 \textwidth]{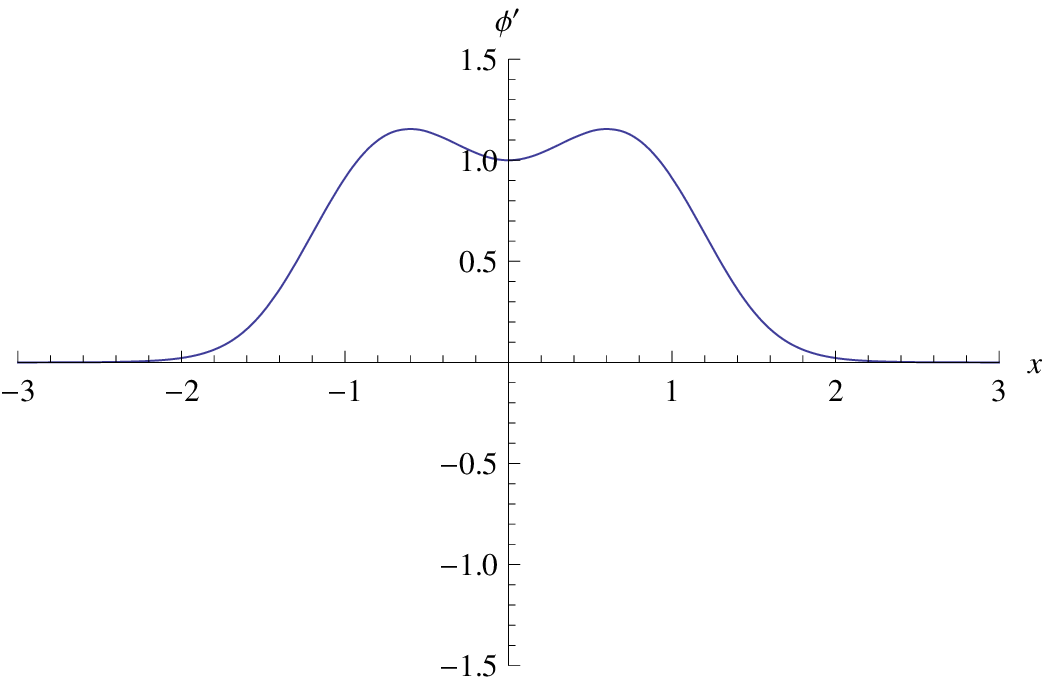}
\caption{The kink interpolating between $-\sqrt{2}$ and $\sqrt{2}$ and its first derivative.} \label{fig4}
\end{figure}

\subsection{The models of Bazeia, Menezes and Petrov (BMP)}
Now let us go back to the building blocks (\ref{building-b}) and choose linear combinations such that all higher than second powers in the auxiliary field $F$ cancel. The right choice is
\begin{eqnarray}
( {\cal L}^{(k)}_{\rm BMP})_{\psi =0} &\equiv & 
\sum_{i=0}^{k-1} \binom{k-1}{i}(-1)^i ( {\cal L}^{(k-i,i)})_{\psi =0} \nonumber \\ 
&=&  \left( (\partial^\mu\phi\partial_\mu\phi) + F^{2} \right) (\partial^\nu \phi \partial_ \nu \phi )^{k-1} 
\label{k-n-sum2}
\end{eqnarray}
and linear combinations of these lead to
\begin{equation} \label{BMP-new}
 {\cal L}_{\rm BMP} = \sum_{k=1}^\infty \beta_k {\cal L}^{(k)}_{\rm BMP} = (F^2 + \partial_\mu \phi \partial^\mu \phi ) \sum_k \beta_k (\partial_\mu \phi \partial^\mu \phi )^{k-1} \equiv (F^2 + \partial_\mu \phi \partial^\mu \phi ) f(\partial_\mu\phi \partial^\mu \phi) .
\end{equation}
This bosonic Lagrangian is identical to the bosonic sector of the original Lagrangian of  BMP \cite{bazeia2} which is based on the superfield
\begin{equation} \label{chi}
\partial_\mu \Phi \partial^\mu \Phi = \partial_\mu \phi \partial^\mu \phi + 2 \theta^\alpha \partial_\mu \phi \partial^\mu \psi_\alpha -  2\theta^2 \partial_\mu \phi \partial^\mu F -\theta^2 \partial_\mu \psi^\alpha \partial^\mu \psi_\alpha .
\end{equation}
Indeed, the bosonic component of the superfield $D_\alpha \Phi D^\alpha \Phi$ only consists of a term proportional to $\theta^2$, therefore multiplying this superfield by an arbitrary function of the above superfield (\ref{chi}), $f(\partial_\mu \Phi \partial^\mu \Phi)$, only the theta independent term $f(\partial_\mu \phi \partial^\mu \phi )$ will contribute, leading to the Lagrangian
\begin{equation} \label{Baz}
{\cal L}_{\rm BMP} =  [f(\partial_\mu \Phi \partial^\mu \Phi )\frac{1}{2} D_\alpha \Phi D^\alpha \Phi ] _{\theta^2 ,\psi =0} 
= f(\partial_\mu \phi \partial^\mu \phi )(F^2 + \partial_\mu \phi \partial^\mu \phi )
\end{equation}
which obviously coincides with Eq. (\ref{BMP-new}). The fermionic sectors, on the other hand, will in general not coincide.  Adding a potential term in order to allow for kink solutions, one gets as usual
\begin{equation}
{\cal L}_{\rm BMP}^{(P)} = f(\partial_\mu \phi \partial^\mu \phi ) (F^2 + \partial_\mu \phi \partial^\mu \phi ) - P' (\phi )F
\end{equation}
which after eliminating the auxiliary field $F$ using its algebraic field equation
\begin{equation}
F= \frac{P'}{2f}
\end{equation}
becomes
\begin{equation}
{\cal L}_{\rm BMP}^{(P)} = f (\frac{P'^2}{4f^2} + \partial_\mu \phi \partial^\mu \phi ) -
\frac{P'^2 }{2f} = \partial_\mu \phi \partial^\mu \phi f - \frac{P'^2}{4f}
\end{equation}
where we suppressed the arguments of $P$ and $f$ in the last expression to improve readability.
The once integrated static field equation (\ref{zero-press}), after a simple calculation, leads to 
\begin{equation}
\frac{1}{4f^2}(f+2Xf_{,X}) (8Xf^2 + P'^2)=0
\end{equation}
where now $X=-\frac{1}{2}\phi'^2$ and $f=f(-\phi'^2)$ and, therefore, to the two equations
\begin{equation} \label{BPS-eq}
2\phi ' (x)f(-\phi '(x)^2)= \pm P' (\phi) .
\end{equation}
For some choices of $f$ and $P$ these equations lead to soliton solutions, where some explicit examples have been given in \cite{bazeia2} and in \cite{twins}.
Here we want to focus, instead, on the BPS property of the model.
The energy functional for static configurations may, in fact, be re-written in a BPS form \cite{bazeia2}, exactly like for the standard supersymmetric scalar field theory, from which both the first order equations and the energy bound follow immediately. Indeed, the energy functional may be written like
\begin{equation}
{ E}_{\rm BMP}^{(P)} = \int dx \left( \phi '^2 f + \frac{P'^2}{4f}\right) = \int dx \left( \frac{1}{4f}(2\phi ' f \mp P')^2 \pm \phi ' P' \right)
\end{equation}
and for a solution to the first order (or BPS) equation (\ref{BPS-eq}) (we take the plus sign for definiteness) the resulting energy is therefore 
\begin{equation}
{ E}_{\rm BMP}^{(P)} = \int_{-\infty}^\infty dx \phi ' P' = \int_{\phi (-\infty)}^{\phi (\infty)} d\phi P' =P(\phi (\infty)) - P(\phi (-\infty)).
\end{equation}

\section{Supersymmetric baby Skyrme models}
Here we want to briefly describe the SUSY baby Skyrme models in 2+1 dimensions which were first found and investigated in \cite{bS}.
The non-supersymmetric baby Skyrme models are given by the class of Lagrangians
\begin{equation} \label{BS-lag}
L= \frac{\lambda_2}{2} L_2 + \frac{\lambda_4}{4} L_4 + \lambda_0 L_0
\end{equation}
where the $\lambda_i$ are coupling constants and the $L_i$ are (the subindices refer to the number of derivatives)
\begin{equation} 
L_2 = \partial_\mu \vec \phi \cdot \partial^\mu \vec \phi 
\end{equation}
(the standard nonlinear sigma model term),
\begin{equation}
L_4 = -(\partial_\mu\vec{\phi}\times\partial_\nu\vec{\phi})^2
\end{equation}
(the Skyrme term)
and a potential term
\begin{equation}
L_0 = -V(\phi_3)
\end{equation}
which is usually assumed to depend only on the third component $\phi_3$ of the field. The three-component field vector $\vec \phi$ obeys the constraint $\vec \phi^2 =1$. 

The problem now consists in finding the supersymmetric extensions ${\cal L}_i$ of all the contributions $L_i$ to the Lagrangian (\ref{BS-lag}). For the non-linear O(3) sigma model term $L_2$ this supersymmetric extension was found long ago in \cite{witten1}, \cite{alvarez}. One simply chooses the standard SUSY kinetic term $ \frac{1}{2}(D^\alpha \Phi^i D_\alpha \Phi^i)_{\theta^2}$ for the lagrangian ${\cal L}_2$ and imposes the constraint $\vec \phi^2 =1$ on the superfield, i.e., $\vec \Phi^2 =1 $, which in components reads
\begin{eqnarray}
\phi^i \cdot\phi^i &=&1\\
\phi^i \cdot \psi ^i_\alpha&=&0\\
\phi^i \cdot F^i &=&\frac{1}{2}\bar{\psi}^a\psi^a .
\end{eqnarray}
It may be checked easily that the constraint $\vec \Phi^2 =1$ is invariant under the $N=1$ SUSY transformations
\begin{equation}
\delta \phi^i =  \epsilon^\alpha \psi^i_\alpha \; , \quad \delta \psi^i_\alpha = -i\partial_\alpha{}^\beta \epsilon_\beta \phi^i - 
\epsilon_\alpha F^i \; , \quad 
\delta F^i = i  \epsilon^\beta  \partial_\beta{}^\alpha \psi^i_\alpha .
\end{equation}
Like in Section 2, we shall be interested mainly in the bosonic sectors of the SUSY extensions, therefore we set $\psi_\alpha^i =0$ in the following. For $\psi =0$, the constraints simplify to
\begin{eqnarray}
\phi^i\cdot\phi^i&=&1 \label{bos-constr-1} \\
\phi^i \cdot F^i &=&0. \label{bos-constr-2} 
\end{eqnarray}  
 For the SUSY extension of $L_2$ we get
\begin{equation}
({\cal L}_2)_{\psi =0} = \frac{1}{2}[ (D^\alpha \Phi^i D_\alpha \Phi^i)]_{\theta^2, \psi =0} =  F^iF^i + \partial^\mu \phi^i \partial_\mu \phi^i  ,
\end{equation}
whereas for the Skyrme term we choose the same linear combination of the building blocks (\ref{building-b}) such that mixed terms coupling auxiliary fields $F^i$ and derivatives of the scalar fields $\partial_\mu \phi^i$ are absent,
\begin{eqnarray}
( {\cal L}_4)_{\psi =0} &=&  \frac{1}{2}\epsilon_{ijk}\epsilon_{i'j'k}[ ( D^\alpha \Phi^i D_\alpha \Phi^{i'} D^2 \Phi^j D^2 \Phi^{j'} +
D^\alpha \Phi^j D_\alpha \Phi^{j'} D^2 \Phi^i D^2 \Phi^{i'}  )]_{\theta^2 ,\psi =0} \nonumber \\
&& -\; \frac{1}{8}\epsilon_{ijk}\epsilon_{i'j'k}[ ( D^\alpha \Phi^i D_\alpha \Phi^{i'} D^\gamma D^\beta \Phi^j D_\gamma D_\beta \Phi^{j'} +
\nonumber \\
&& \qquad \qquad \qquad
+\;  D^\alpha \Phi^j D_\alpha \Phi^{j'} D^\gamma D^\beta \Phi^i D_\gamma D_\beta \Phi^{i'}
)]_{\theta^2 ,\psi =0} \nonumber \\
&=& \epsilon_{ijk}\epsilon_{i'j'k} (F^i F^{i'} F^j F^{j'} - \partial_\mu \phi^i \partial^\mu \phi^{i'} \partial_\nu \phi^j \partial^\nu \phi^{j'} ) = 
-(\partial_\mu \vec \phi \times \partial_\nu \vec \phi )^2
\end{eqnarray}
and it turns out that due to the antisymmetry of the Skyrme term the contribution quartic in $F^i$ is identically zero. Finally, 
for the potential term we choose, as usual 
\begin{equation}
({\cal L}_0)_{\psi =0} = [ P(\Phi_3) ]_{\theta^2 ,\psi =0} = - F_3 P' (\phi_3) ,
\end{equation}
where $P$ is the superpotential and the prime denotes derivation w.r.t. its argument $\phi_3$.
The resulting bosonic lagrangian is
\begin{eqnarray}
({\cal L})_{\psi =0} &=&  \frac{\lambda_2}{2} [(\vec F)^2 + \partial_\mu \vec \phi \cdot \partial^\mu \vec \phi ] 
 - \; \frac{\lambda_4}{4} (\partial_\mu \vec \phi \times \partial_\nu \vec \phi )^2 \nonumber \\ &&
- \lambda_0 F_3 P' + \mu_F (\vec F \cdot \vec \phi )
+ \mu_\phi (\vec \phi^2 -1) 
\end{eqnarray}
where $\mu_F$ and $\mu_\phi$ are Lagrange multipliers enforcing the constraints (\ref{bos-constr-2}) and (\ref{bos-constr-1}).

The (algebraic) field equation for the field $\vec F$ is 
\begin{equation} \label{F-eq-bS}
\lambda_2  F^i - \lambda_0 \delta^{i3} P' (\phi_3) + \mu_F \phi^i =0.
\end{equation}
Multiplying by $\vec \phi$ we find for the Lagrange multiplier 
\begin{equation}
\mu_F = \lambda_0 \phi_3 P' 
\end{equation}
and for the auxiliary field $\vec F$
\begin{equation}
F^i = -\frac{\lambda_0}{\lambda_2} (\phi_3 \phi^i - \delta^{i3})P'
\end{equation}
and, therefore, for the bosonic Lagrangian
\begin{eqnarray}
({\cal L})_{\psi =0} &=& \frac{\lambda_2}{2} [(\vec F)^2 + \partial_\mu \vec \phi \cdot \partial^\mu \vec \phi ] 
 - \frac{\lambda_4}{4} (\partial_\mu \vec \phi \times \partial_\nu \vec \phi )^2 + \lambda_0 F_3 P' 
+ \mu_\phi (\vec \phi^2 -1) \nonumber \\
&=& \frac{\lambda_2}{2}  \partial_\mu \vec \phi \cdot \partial^\mu \vec \phi 
 - \frac{\lambda_4}{4} (\partial_\mu \vec \phi \times \partial_\nu \vec \phi )^2 - \frac{\lambda_0^2}{2\lambda_2}(1-\phi_3^2)P'^2 
+ \mu_\phi (\vec \phi^2 -1) .
\end{eqnarray}
This is exactly the standard (non-supersymmetric) baby Skyrme model with a potential term given by
\begin{equation}
V(\phi_3) = \frac{\lambda_0}{2\lambda_2} (1-\phi_3^2) P'^2 (\phi_3) .
\end{equation}
Obviously, all positive semi-definite potentials $V(\phi_3)$ may be obtained by an appropriate choice for the superpotential $P(\phi_3)$.

{\em Remark:} the relation between superpotential and potential differs slightly (by the additional factor $(1-\phi_3^2)$) from the standard SUSY relation, due to the constrained nature of the superfield $\vec \Phi$.   

{\em Remark:} The baby Skyrme model has a Bogomolny bound in terms of the topological charge (winding number) of the scalar field $\vec \phi$, but nontrivial solutions in general do not saturate this bound. There exist, however, two limiting cases where nontrivial solutions do saturate a Bogomolny bound and solve the corresponding  first order Bogomolny equations. One might wonder whether these limiting cases allow for the supersymmetric extension discussed in this section, as well. The first limiting case is the case of the pure O(3) sigma model where both the potential and the quartic (Skyrme) term are absent, and, as discussed above, it is well-known that this case has a supersymmetric extension. 
Concerning the second case, it has been found recently that the model without the quadratic O(3) sigma model term (i.e., $\lambda_2 =0$) has a Bogomolny bound and nontrivial BPS solutions saturating this bound. Further, this restricted model has an infinite number of symmetries and conservation laws \cite{comp-bS}. Given the close relation between Bogomolny solutions and supersymmetry, one might expect that this limiting case should have the supersymmetric extension, too. Surprisingly, this is not true. The field equation (\ref{F-eq-bS}) for $\vec F$ for the case $\lambda_2 =0$ reads 
$$ \lambda_0 \delta^{i3} P' (\phi_3) + \mu_F \phi^i =0.$$
It does not contain $\vec F$ at all, so $\vec F$ itself is a Lagrange multiplier in this case. For a nontrivial field configuration $\vec\phi$, the only solution of this equation is $\mu_F =0$ and $\lambda_0=0$, therefore the potential term is absent. 
We conclude that, although the model consisting only of the quartic Skyrme term ${\cal L}_4$ does allow for a supersymmetric extension (but does not support soliton solutions), the model consisting of both the quartic Skyrme term and the potential term ${\cal L}_0$ does not allow for the supersymmetric extensions discussed in this section. 

\section{Discussion}
It was the main purpose of this contribution to give a brief overview of recent results about the SUSY extensions of scalar field theories with non-standard kinetic terms (K field theories), where the methods to obtain the SUSY extensions described in this article are directly applicable in 1+1 and 2+1 dimensions, due to the similarity of the SUSY representations in these two cases. Concretely, in 1+1 dimensions we described a class of SUSY scalar field theories \cite{susy2} which, in the bosonic sector, may still be written as the sum of a generalized kinetic term and a potential. For these theories, the vacuum structure of the potential is crucial for the existence of topological kink solutions, similar to the case of a standard scalar field theory. There remain, nevertheless, some important differences. 
Due to the enhanced nonlinearity of these supersymmetric K field models, the first order field equations have a higher number of roots leading to a larger number of possible kink solutions, and kinks do not necessarily have to connect adjacent vacua. A related issue is the possibility to join different solutions forming additional kinks in the space ${\cal C}^1$ of continuous functions with a continuous first 
derivative. Another main difference is the fact that these generalized SUSY models do not have the BPS property, in spite of the fact that they support topological solitons and that the corresponding static field equations may be written in a first order form (i.e. they have a first integral of motion).
The models introduced in \cite{bazeia2} (and briefly discussed in Section 2.3), on the other hand, not only have static first order equations and the corresponding kink solutions, they also have the BPS property. This seems to be related to the fact that in these models the auxiliary field still shows up at most quadratically, like in the standard case. On the other hand, the models of \cite{bazeia2} cannot be expressed as a generalized kinetic term plus a potential, because the auxiliary field couples to derivatives. 

As a specific application in 2+1 dimensions, we briefly described the construction of $N=1$ SUSY extensions of baby Skyrme models originally found in \cite{bS}, which works for arbitrary non-negative potentials. It is wellknown that nontrivial soliton solutions of baby Skyrme models are not of the BPS type, in spite of the existence of a Bogomolny bound in these models, so the issue of the BPS propery of the SUSY extensions does not show up directly. It came as a surprise, however, that precisely those restricted baby Skyrme models which do have nontrivial BPS solutions \cite{comp-bS} {\em cannot} be supersymmetrically extended by the methods of \cite{bS}. All these findings seem to indicate that the relation between the existence of topological soliton solutions and the BPS property is less direct or more involved in SUSY K field theories than it is in standard SUSY field theories. 

A related question is whether in SUSY K field theories there exists a relation between the existence of topological solitons with their topological charges, on the one hand, and central extensions in the SUSY algebra in the soliton background, on the other hand, as is known to hold in the standard case \cite{witten-olive}. An investigation of this issue obviously requires the inclusion of fermions, i.e., the study of the full SUSY K field theories, not just of their bosonic sector. The inclusion of fermions in the SUSY Lagrangians described in the present contribution does not present fundamental obstacles, the only practical difficulty being that, for purely combinatorical reasons, the resulting expressions will be rather lengthy.  In any case, the inclusion of fermions and the study of the full SUSY algebra is the next obvious step in the further investigation of these theories.

Another question of some interest which may be posed specifically for the SUSY extension of the baby Skyrme model is like follows. It is wellknown that the non-SUSY baby Skyrme model has another, classically equivalent formulation in terms of a CP$^1$ field instead of the O(3) invariant field $\vec \phi$. The simplest way to arrive at this CP$^1$ version is via a stereographic projection from the two-sphere spanned by $\vec \phi$ to the complex plane spanned by a complex field $u$. The obvious question to ask is whether the CP$^1$ version of the baby Skyrme model has a SUSY extension, as well, and if this is the case, whether the O(3) SUSY extension and the CP$^1$ SUSY extension are equivalent. It is known that the answer to both questions is positive in the restricted case where both the Skyrme term and the potential are absent, i.e., in the so-called O(3) nonlinear sigma (or CP$^1$) model. Despite their equivalence, the way the SUSY extensions are constructed is quite different, however. In the O(3) formulation of the nonlinear sigma model, the three components $\vec \phi$ of the O(3) field are promoted to real scalar superfields just as described in Section 3 \cite{witten1}, so only an $N=1$ SUSY is manifest, although the model eventually has an extended $N=2$ SUSY. In the CP$^1$ formulation of the nonlinear sigma model, on the other hand, one uses the fact that the non-SUSY Lagrangian is the pullback of a target space Kaehler metric. The complex CP$^1$ field $u$ is therefore promoted to a complex chiral superfield and the theory has manifest $N=2$ SUSY by construction \cite{zumino}. We found, however, only a $N=1$ SUSY extension of the baby Skyrme model, therefore a complex chiral superfield with $N=2$ extended SUSY in the CP$^1$ formulation is probably not a good starting point. It may be that a manifestly $N=1$ SUSY version of the CP$^1$ nonlinear sigma model must be found before one can try to answer the question posed above. This issue will be investigated further, as well.  

As far as possible applications are concerned, the natural arena seems to be the area of cosmology. Indeed, if scalar field theories like the ones discussed in the present contribution are interpreted as effective theories which derive from a more fundamental theory with supersymmetry (like string theory), then it is natural to study the supersymmetric versions of the effective models. Also the presence of non-standard kinetic terms is natural for effective theories. If, in addition, the defect formations and phase transitions (e.g. from a symmetry breaking phase to a symmetric phase) relevant for cosmological considerations occur at time or energy scales where supersymmetry is still assumed unbroken, then also the defect solutions of the {\em supersymmetric} effective field theories are the relevant ones.

\ack
The authors acknowledge financial support from the Ministry of Science and Investigation, Spain (grant FPA2008-01177), 
the Xunta de Galicia (grant INCITE09.296.035PR and
Conselleria de Educacion), the
Spanish Consolider-Ingenio 2010 Programme CPAN (CSD2007-00042), and FEDER.

\end{document}